\def\year{2020}\relax
\documentclass[letterpaper]{article} % DO NOT CHANGE THIS
\usepackage{aaai20}  % DO NOT CHANGE THIS
\usepackage{times}  % DO NOT CHANGE THIS
\usepackage{helvet} % DO NOT CHANGE THIS
\usepackage{courier}  % DO NOT CHANGE THIS
\usepackage[hyphens]{url}  % DO NOT CHANGE THIS
\usepackage{graphicx} % DO NOT CHANGE THIS
\usepackage{graphicx}  % DO NOT CHANGE THIS
\usepackage{xspace}
\usepackage{amsmath}
\usepackage{comment}
\usepackage{array}
\usepackage{color}
\usepackage{mathtools}
\usepackage{dsfont}
\usepackage{booktabs} % For formal tables

\newcommand{\kenneth}[1]{{\small\color{blue}{\bf\xspace#1 -Kenneth}}}

\newcommand{\system}{KG-Story\xspace}

\title{On Automating Conversations}
\author{
Ting-Hao (Kenneth) Huang\\
College of Information Sciences and Technology\\
Pennsylvania State University, University Park, PA, USA\\
\texttt{txh710@psu.edu}\\
}

\usepackage{enumitem}

\newcommand{\chorus}{Chorus\xspace}
\newcommand{\ic}{InstructableCrowd\xspace}
\newcommand{\ga}{Guardian\xspace}

\newcommand{\ev}{Evorus\xspace}
\newcommand{\ig}{Ignition\xspace}

\begin{document}
\maketitle

\begin{abstract}

From 2016 to 2018, we developed and deployed Chorus, a system that blends real-time human computation with artificial intelligence (AI) and has real-world, open conversations with users. We took a top-down approach that started with a working crowd-powered system, \chorus, and then created a framework, \ev, that enables \chorus to automate itself over time. Over our two-year deployment, more than 420 users talked with \chorus, having over 2,200 conversation sessions. This line of work demonstrated how a crowd-powered conversational assistant can be automated over time, and more importantly, how such a system can be deployed to talk with real users to help them with their everyday tasks.
This position paper discusses two sets of challenges that we explored during the development and deployment of \chorus and \ev: the challenges that come from being an ``agent'' and those that arise from the subset of conversations that are more difficult to automate.

\end{abstract}

\section{Deploying and Automating Chorus}

\begin{comment}

\begin{figure}[t]
  \centering
  %\includegraphics[width=0.99\columnwidth]{img/evorus_overview.png}
  \includegraphics[width=0.99\columnwidth]{figure/evorus-framework-overview.png}
  %\vspace{-.3pc}
  \caption{\ev is a crowd-powered conversational assistant that automates itself over time by {\em (i)} learning to include responses from chatterbots and task-oriented dialog systems over time, {\em (ii)} reusing past responses, and {\em (iii)} gradually reducing the crowd's role in choosing high-quality responses by partially automating voting.}
  %\vspace{-1pc}
  \label{fig:ev:system-overview}
\end{figure}
\end{comment}

Interaction in rich natural language enables people to exchange ideas efficiently and achieve their goals more quickly. Modern personal intelligent assistants, such as Apple's Siri and Amazon's Echo, utilize conversation as their primary communication channel and illustrate a future where conversing with computers will be as easy as talking to a friend. However, despite decades of research, modern conversational assistants are still limited in domain, expressiveness, and robustness. 

\begin{figure*}[t]
  \centering
  \includegraphics[width=0.99\textwidth]{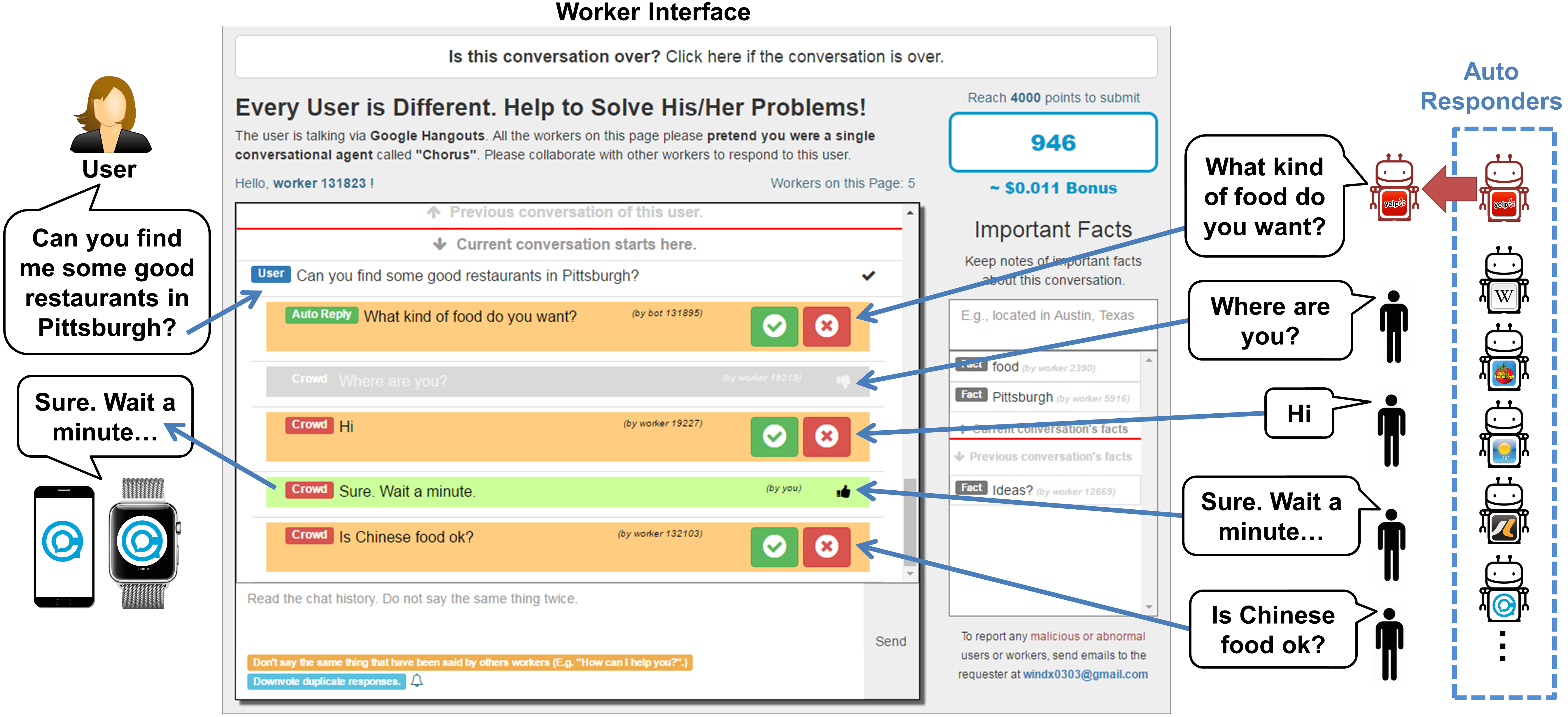}
  %\vspace{-0.5pc}
  \caption{The \ev worker interface allows workers to propose responses and up/down vote candidate responses. The up/down votes give \ev labels to use to train its machine-learning system to automatically gauge the quality of responses. \ev automatically expires response candidates upon acceptance of another in order to prevent workers from voting through candidate responses that are good but no longer relevant. Workers can tell each message is sent by the end-user (blue label), a worker (red label), or a chatbot (green label) by the colored labels.}
  %We made the source transparent because we observed workers would other report the occasional odd responses by automated systems.}
  \label{fig:ev:ui-system}
  %\vspace{-1pc}
\end{figure*}

In response, we created a system that blends real-time human computation with AI and has real-world, open conversations with users. Instead of bootstrapping automation from the bottom up with only automatic components, we took a top-down approach that started with a working crowd-powered system. We developed and deployed a crowd-powered conversational assistant, {\em \chorus}~\cite{chorusDeploy}, and then created a framework, \ev, that enables \chorus to automate itself over time~\cite{Evorus}. \ev allowed external task-oriented chatbots and chatterbots to be added into \chorus to take over parts of conversations, reuse crowd-submitted responses to answer future similar questions, and gradually learn to select high-quality responses to reduce its reliance on crowd oversight. Figure~\ref{fig:ev:ui-system} shows the worker interface and a system overview of \ev. To make the deployment more robust, we invented a new recruiting method, the Ignition model~\cite{ignition}, to hire workers quickly. In our two-year deployment, more than 420 users talked with \chorus, having over 2,200 conversation sessions.

During the deployment of \ev, conversations averaged 9.90 messages sent from users, 13.58 messages sent from crowd workers, and 1.93 messages sent from automatic chatterbots~\cite{Evorus}. Thus, \textbf{automated responses were chosen 12.44\% of the time}. As a comparison, without any automation, a conversation averaged 8.73 user messages and 12.98 crowd messages. To examine conversation quality, we sampled conversations with accepted automatic responses and a matching set without automated contributions. For each, 8 MTurk workers rated [Satisfaction, Clarity, Responsiveness, Comfort], which was based on the PARADISE's objectives for evaluating dialogue performance~\cite{paradise} and the Quality of Communication Experience metric~\cite{liu2010quality}, on a 5-point Likert scale (with 5 being the best.) The original conversations (N=46) had an average rating of [3.47, 4.04, 3.88, 3.56], while those with automatic responses (N=54) had scores of [3.57, 3.74, 3.52, 3.66]. The similar results suggest that the automatic components did not make conversations worse. 

Having human workers in the system loop enabled \chorus to talk with real users to help them with their everyday tasks and demonstrated to us the challenges of conversational assistants that have not yet been fully explored by state-of-the-art automated chatbots.
In this position paper, we discuss two sets of challenges that are less discussed in the AI community~\cite{huang_dissertation_2018}: the common drawbacks of being an ``agent'' and human conversations beyond the scope of task-oriented dialogues.

%----------------------------

%In this position paper, we discuss the impact of automation in the deployed Chorus system, the power and the limitation of the ``agent'' metaphor, and the future of automating conversations.
%\kenneth{talk about results of Evorus}

\section{Being an ``Agent''}
%\kenneth{how and why this is relevant to AI and Work? A thinking framework that helps to make the trade-offs more explicit.}
\chorus, \ev, and the concept of a ``crowd agent''\cite{crowd-agency} strongly imply the interaction metaphor of an ``agent.'' We are aware of the historical debate about the pros and cons of the ``agent'' metaphor for human-computer interaction. 
Ben Shneiderman famously opposes this model. 
%As early as 1986, 
He argued that people are different from computers, and ``human-human interaction is not necessarily an appropriate model for human operation of computers.''~\cite{shneiderman2010designing} In 1997, Shneiderman and Pattie Maes had a landmark debate about the agent metaphor~\cite{Shneiderman:1997:DMV:267505.267514} and the pros and cons of ``direct manipulation'' and ``software agents.'' Shneiderman argued for the power of good user interfaces and visualizations and again stated that human-human interaction should not be the model for human operation of computers. Maes argued that having full control of unlimited actionable items and overly complex tasks is impractical, and thus future users will need ``extra eyes and extra ears'' to which to delegate tasks. In 2017, they revisited this debate. In this latest round, Maes argued for the concept of ``human-computer integration,'' while Shneiderman countered that computers are tools rather than equal partners of humans, and the goal of interactive technology should be ``ensuring human control, while increasing the level of automation.''~\cite{Farooq:2017:HCI:3027063.3051137}

We would like to start our discussion with an example of a real-world complex task: tax preparation. To prepare their taxes, most people either hire human experts ({\em i.e.,} accountants or tax professionals) or use specialty tax software ({\em e.g.,} TurboTax). These reflect the two sides of the ``agent'' metaphor debate: Human experts are hired agents who use conversations and natural languages ({\em e.g.,} email) as primary means of communication; conversely, tax programs use a nicely designed workflow and user interface to transform the complex tax preparation process into a task that most users can accomplish. Each solution has its advantages and disadvantages. Human experts are more intelligent at understanding user needs and preferences, more flexible in communication, better at acknowledging context to provide customized suggestions, and able to save time and effort for users. However, the gaps between a human professional's capability and what state-of-the-art technologies can do is significant. Without proper automated solutions, hiring a tax professional is expensive, and an expert's time is not scalable. On the other hand, tax software gives users more direct control and instant feedback on the interface and is typically more affordable. However, software is less flexible, harder to customize for complex tax situations, and most importantly, requires much more time and effort from users, who must manually input data. 

Our work aimed to relax the limitations set by hiring humans and attempted to reveal the true potential of the agent metaphor. The gap between what people want to accomplish (for example, interacting naturally) and what technology is capable of -- known as the ``social-technical gap'' -- is one of the central challenges for studying computer-supported cooperative work (CSCW) and HCI~\cite{ackerman2000intellectual}. Our work showed how to use automation to reduce the gaps between human agents and automated solutions, but to fully explore the real trade-offs of agents as compared to direct manipulation (tools), we need more competent systems. We agreed with Maes's insight that the software agent is powerful. However, we are also aware that to become as useful as we imagine agents could be, we must be much closer to human-level capabilities.

\paragraph{Miscommunication}
Users need to communicate problems and context to an agent to delegate tasks, which provides opportunities for miscommunication.
Although \chorus enabled open conversation with users, the flexibility and expressiveness of natural language mean ambiguity is inevitable in human languages.
Therefore, users' verbal specifications could be interpreted in different ways. 
Misinterpretation of language was not the only cause of miscommunication: Users also omitted important contextual information until prompted by the agent. 
The agent cannot know beforehand all the context, preferences, and constraints of the user, so miscommunication is inevitable. 
%It is impossible for the agent to know beforehand all the context, preferences, and constraints of the user, so miscommunication is inevitable. 
However, one of the primary advantages of our system is its ability to adapt to the user through open, multi-turn communication.

\paragraph{Hard or Expensive for Hands-on Controls}
Authorizing an agent to complete tasks means users do not have hands-on control of all task details. The agent often needs to make small decisions without constantly checking with the user, as such interruptions would devalue the automated component of the system. However, lack of control could be a problem for tasks that need to be precisely customized or when users simply prefer full control. In other words, if users hire an agent to personalize something for them ({\em e.g.,} a gift for mom), it would be difficult or expensive for the agent to mimic the same level of personalization that the user could provide. This is one of the fundamental limitations when hiring any agency.

\paragraph{Potentially Longer Task Completion Time}
Another limitation of outsourcing a task to an agent is that the end-to-end task completion time could be longer than that required if the users did the task themselves, especially for tasks where users were proficient. Users sometimes did not use \chorus due to the task being difficult. For tasks that were familiar or easy, it might take \chorus longer than the user because of the need to communicate with the agent, iterate back and forth to narrow down results, and finally obtain a result. In our experiments with \ic~\cite{huang2019instructablecrowd}, in some scenarios, the time the system took to produce IF-THEN rules was longer than that required by users on a mobile interface. While the users do not need to commit any time or effort while waiting, we acknowledge that communicating with \chorus takes time and the end-to-end completion time could be longer.

\subsubsection{Higher User Expectation}
Finally, another fundamental limitation is that anthropomorphic agents create higher expectations of users, which can harm the user experience or create greater frustration when problems arise. For instance, a user might become more frustrated with an Amazon Echo device that failed to execute a command to ``turn off the light'' than with a manual switch that did not work. However, we also argue that higher user expectations will motivate users to explore new tasks that they did not think were possible and thus push the boundaries of the capability of conversational assistants. While the system ended up disappointing some users~\cite{nytimes-siri-disappointment}, it opened up many new research problems and use cases, influencing how today's users interact with personal assistants. Over the long term, higher expectations are not a limitation but an opportunity.

\section{The Conversations That Are\\Harder to Automate}
%\kenneth{1. what is hard for current AI systems? 2. But why we should do them? from business point of view, maybe we just don't need to do them?}
While having human workers in the loop enables \chorus to hold realistic conversations with users, we observed limitations in terms of handling or automating social conversation and open-ended discussions. These limitations reflect the current bottleneck of conversational assistants and could indicate the future challenges of automating conversations.

\subsubsection{Automating Domain-Independent Tasks and Social Conversations}

The \ev framework focused only on automating domain-specific tasks.
However, some real-world tasks are cross-domain or even domain-independent.
It would be difficult for \ev to automate these tasks because its bot-selection algorithm used topic similarity to choose chatbots. For instance, negotiation ({\em e.g.,} salary or price) is a common task in the real world that can happen in many different domains, including real estate, shopping, or trading.
If \ev included a ``negotiation bot'' that could negotiate with users, the bot-selection algorithm would have a hard time using domain similarity to decide when to use it.
To date, our work has focused more on tasks than on social conversations.
While the current version of \ev included four chatbots to provide general responses, we did not incorporate approaches or knowledge about social conversations in our system.
As in non-domain tasks, the bot-selection algorithm is also inefficient in selecting social chatbots because they do not represent a clear, single domain.

\subsubsection{Open-Ended Discussions}
The deployed version of \chorus did not help workers organize information in a structured way, something that would help support open-ended discussions. The system only had a ``memory board'' where workers could write down a \textit{list} of important facts. While this approach was sufficient to support most conversations, a list is insufficient for organizing all the arguments and information for long, open-ended long discussions such as ``which school should I go to?'' or ``what's the meaning of working?''

Ackerman studied how computer systems can be used to enhance organizational or group memory~\cite{Ackerman:1998:AOM:290159.290160,Ackerman:1990:AGT:91474.91485}. Lasecki and Bigham experimented using workers' ratings to rank notes recorded on \chorus' memory board according to their relevance to the current conversation~\cite{lasecki2013automated}. 
Gouravajhala {\em et al.} created a more advanced version of a memory board, Mnemo, where workers can summarize a note and estimate its longevity by answering questions like ``Will this fact still be true by [the end of today/next week/next month/next year]?''~\cite{gouravajhalafinding}. However, we did not focus on improving the memory board to enable more advanced structures for organizing information.

Furthermore, \chorus and \ev used worker consensus to choose high-quality responses, assuming that most workers likely agreed on the same high-quality responses. However, sometimes this assumption did not hold. In realistic open-ended discussions, topics are often subjective, where workers do not agree with each other easily or quickly. \chorus currently cannot hold these types of conversations without breaking the single-agent illusion. One potential solution would be to automatically detect such topics and alter the worker interface or voting mechanism correspondingly.

\section{The ``Top-Down'' Research Approach}
One lesson of our work is the effectiveness of the ``top-down'' research approach. %We started with a working, deployed system, learned to improve it, and created a framework that allows the system to automate itself over time.
One core advantage of this top-down approach is that users can start using the system on day one, providing realistic data to guide future automation. 
Furthermore, thanks to the oversight of crowd workers, such human-in-the-loop frameworks allow automated components to make more mistakes, opening more spaces for algorithms to try different strategies.

At a higher level, crowd-powered approaches are robust, intelligent, and do not require training data, but they are slow and costly. In contrast, automated approaches are fast, scalable, and more affordable, but they require large amounts of training data and have many more capability limitations. Our work attempted to answer this critical question: How do we combine crowd-powered and automated components wisely to leverage the advantages of each while minimizing the disadvantages? Our take was slightly different from the approaches of many prior attempts, which added pieces of human work into a larger automated system architecture. Instead, we introduced pieces of automation into human-powered systems. We believe that -- even in this era of AI, machine learning, and deep learning -- combining human intelligence with automated approaches will result in systems that are more robust, usable, and scalable, thereby creating a better world.

\bibliographystyle{aaai}
\bibliography{aaai.bib}

\end{document}